\documentclass[12pt,a4paper]{article}
\usepackage{graphicx}
\begin{document}
\begin{center}

 {\bf Beyond mean-field description of break-up, transfer and fusion.
}

 \medskip
D. Lacroix$^{1}$, M. Assi\'e$^{1,2}$, S. Ayik$^{3}$, G. Hupin$^{1}$, J.A. Scarpaci$^{2}$ and K. Washiyama$^{1}$

 \smallskip
{\it
 $^{1}$GANIL, CEA and IN2P3, Bo\^ite Postale 5027, 14076 Caen Cedex, France \\

 $^{2}$Institut de Physique Nucl\'eaire, Universit\'e Paris-Sud-11-CNRS/IN2P3, 91406 Orsay, France\\

 $^{3}$Physics Department, Tennessee Technological University, Cookeville, Tennessee 38505, USA. \\
}

\begin{abstract}
Microscopic theories beyond mean-field are developed to include pairing, in-medium nucleon-nucleon collisions as well as effects of 
initial fluctuations of one-body observables on nuclear dynamics. These theories are applied to nuclear reactions. The role
of pairing on nuclear break-up is discussed. By including the effect of zero point motion of collective variables through 
a stochastic mean-field theory, not only average evolution of one-body observables are properly described but also fluctuations. Diffusion coefficients in fusion as well as mass distributions in transfer reactions are estimated. 
\end{abstract}
\end{center}

\bigskip

\section{Introduction}

Time-Dependent Energy Density Functional (TD-EDF) provides a suitable microscopic framework to treat 
both nuclear structure and reactions along the nuclear chart [1,2] . 
In most of current applications, an effective interaction (of Skyrme or Gogny type) is introduced 
to provide an energy functional, denoted ${\cal E}(\rho)$, where $\rho$ is the one-body 
density matrix. 
Then, guided by the Hamiltonian case, equations of motion are        
written in terms of the one-body density evolution 
\begin{eqnarray}
i\hbar \frac{\partial}{\partial t} \rho = [h[\rho],\rho],
\label{eq:puretdhf}
\end{eqnarray}
where 
$h[\rho]\equiv \partial {\cal E}(\rho) /{\partial \rho}$ denotes the mean-field Hamiltonian.
Although this approach is traditionally called Time-Dependent 
Hartree-Fock (TDHF), it should not be 
confused with TDHF derived from a two-body Hamiltonian. At least because of the
parameters of the effective vertex are directly adjusted to experimental observations, EDF 
incorporates much more correlations than a pure Hartree-Fock theory. 
TD-EDF has been improved 
significantly in the past decades and could now be applied without assuming specific 
symmetries in space and includes all terms of the effective interactions used in the static EDF 
[1].

Starting from a pure independent particle state, 
the one-body evolution described by Eq. (\ref{eq:puretdhf}) can be replaced by the evolution of a 
single Slater determinant \footnote{This is due to the fact that $\rho^2 - \rho$ is preserved along the mean-field 
trajectory.}. As a consequence it misses important physical effects which are 
accounted for in "state of the art" EDF dedicated to nuclear structure. These effects are (i) 
Pairing correlations usually treated by considering quasi-particles trial states instead of Slater 
determinants. (ii) Fluctuations of one-body observables and correlations induced by restoration of broken 
symmetries which are generally incorporated using configuration mixing methods, leading to the 
so-called Multi-Reference EDF (MR-EDF) [3-5].  
Due to the numerical effort, Time-Dependent Energy Density Functional devoted to nuclear dynamics  generally 
neglects these correlations and fails to account for the 
richness of phenomena taking place in nuclear dynamics [6]. Recently, we have developed two transport theories 
dedicated to effect (i) and (ii) respectively. Highlights and applications of these new approaches are given below.

\section{Time-Dependent EDF with pairing correlations}

Guided by the Hamiltonian case, different extensions of mean-field
theory have been proposed starting from a generalization of the one-body density equation of motion:
\begin{eqnarray}
i\hbar \frac{\partial \rho}{\partial t} = [h[\rho],\rho] + \frac{1}{2}{\rm Tr}_2 [\tilde v^c_{12}, C_{12}], 
\label{eq:mfcor}
\end{eqnarray} 
where $\tilde v^c_{12}$ denotes the anti-symmetric effective vertex in the correlation channel. 
${\rm Tr}_{2}(.)$ is the partial trace on the second particle while $C_{12} = \rho_{12} -\rho_1 \rho_2 (1-A_{12})$ denotes the two-body correlation matrix defined from the 
two-body density $\rho_{12}$. Here, indices refer to the particle on which the operator is applied while 
$A_{12}$ is the permutation operator. Eq. (\ref{eq:mfcor}) should be complemented by the equation of motion for the two-body correlation.
It is written here as 
\begin{eqnarray}
i\hbar \frac{\partial C_{12}}{\partial t} &=& [h_1[\rho] + h_2[\rho] ,C_{12}]  \nonumber \\
&+& (1-\rho_1)(1-\rho_2){\tilde v}^c_{12}\rho_1 \rho_2 - \rho_1 \rho_2  {\tilde v}^c_{12}(1-\rho_1)(1-\rho_2) ~~ \Leftrightarrow  ~~ 
{B_{12}} \nonumber \\
&+& (1-\rho_1-\rho_2){\tilde v}^c_{12}C_{12}-C_{12}{\tilde v}^c_{12}(1-\rho_1-\rho_2).  \hspace*{1.5cm} \Leftrightarrow  ~~ {P_{12}} \nonumber \\
&+& {\rm Higher~orders}. 
\label{eq:corbph}
\end{eqnarray} 
$B_{12}$ and $P_{12}$ contain correlations associated to in-medium nucleon-nucleon collisions and pairing respectively. Higher-order 
terms (not shown here) contain in particular three-, four-... body effects. When correlations between more than two particles 
are neglected, the resulting theory is known as the Time-Dependent Density-Matrix (TDDM) (see for instance [7]). Even 
in that case, prohibitory numerical efforts are needed due to the explicit treatment of two-body matrices. Guided by the BCS approximation, 
we recently proposed to reduce the complexity by assuming that components of ${\tilde v}^c_{12}$ and $C_{12}$ are non-zero only 
between pairs of time-reversed states.
This approximation, called hereafter TDDM$^{\rm P}$,
leads to important simplifications: (a) the number of correlation
matrix elements to be calculated is significantly
reduced, and (b) $H_{12}$ cancels out and only $B_{12}$ and $P_{12}$ contribute to the correlation evolution.
\begin{figure}[hbtp]
\begin{center}
\includegraphics[width=12cm]{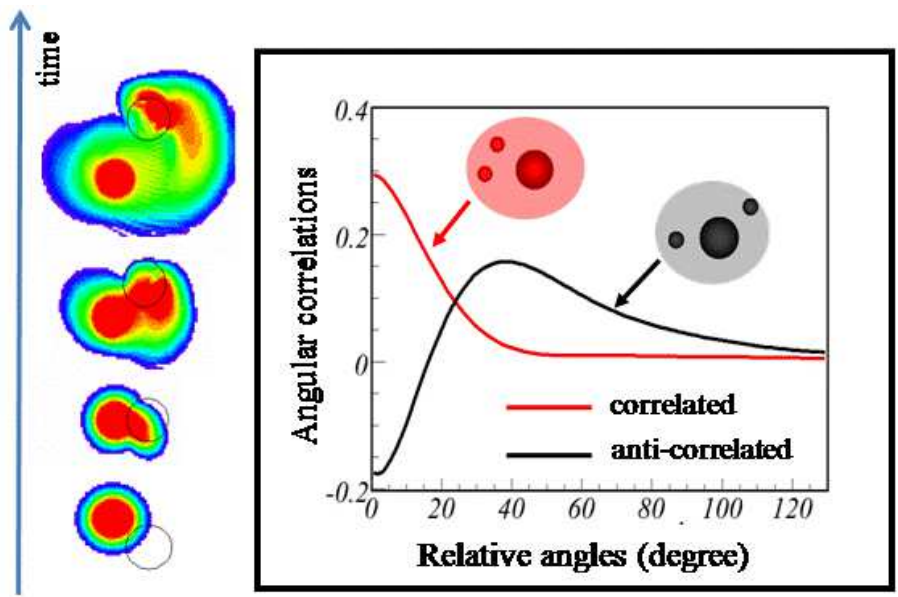}
\caption{(Color online) Left: 
One body density for different times of the dynamical evolution for an $^{16}$O +$^{208}$Pb calculation at 40 A.MeV. The circle 
represents the $^{208}$Pb target. Right: Final distribution of relative angles between emitted nucleons initially correlated (black)
or anti-correlated (red). (Adapted from ref. [8])}
\label{fig1dubna}
\end{center}
\end{figure} 
The TDDM$^{\rm P}$ theory has been recently applied to the nuclear break-up of correlated systems. Figure \ref{fig1dubna} illustrates 
the evolution of the one-body density in a $^{16}$O as a $^{208}$Pb collision partner passes by. The perturbation by the external nuclear 
passing potential induces an emission to the continuum. The one nucleon emission has been extensively studied in ref. [9]. With the TDDM$^P$
theory, one can extend the study of one nucleon case to the two nucleons emission for initially correlated nucleons. In left side 
of figure \ref{fig1dubna}, the relative angle between two nucleons emitted in coincidence has been extracted for initially correlated or anti-correlated systems. As expected intuitively [8,10], in the former case small relative angles are obtained while larger 
relative angles are seen in the latter case. Our study clearly points out that coincidence measurements using nuclear break-up can 
be used as a tool to infer correlation properties inside nuclei.          

\section{Proper treatment of one-body observables fluctuations}

Besides the absence of pairing correlation discussed in previous section, 
mean-field transport theory given by Eq. \ref{eq:puretdhf} provides
a good description of mean values of one-body observables in low energy
reactions. However, it completely fails to describe fluctuations of one-body 
observables. This can directly be traced back to the use of a single Slater 
determinant as a building block. Indeed, as it is well known in the static case, 
the independent particle picture is unable to describe 
properly the zero point motion in collective space pleading for multi-reference approaches 
where several Slater determinants (or quasi-particle states) are considered simultaneously.
   
During the past decades, large efforts have been devoted to develop transport
theories that are able to describe not only mean-values but also fluctuations 
(for a review see [6,11]). However, even 30 years after the first application of TD-EDF, no practical 
solution simple enough to be applied to many different physical phenomena has been really proposed so far.
Recently, we have shown that the theory introduced in ref. [12] and based on stochastic initial values 
in one-body space might provide such a solution. The underlying principle is the following: \\
\noindent
Imagine a correlated system, 
and some one-body observable $Q$ with mean-value $\left\langle  Q \right\rangle$ and fluctuation $\sigma_Q = \langle  Q^2 \rangle - 
\langle  Q \rangle^2$. A single Slater determinant (SD) that minimizes the EDF is generally able 
to reproduce the mean-value but strongly underestimates fluctuations, i.e. 
$\sigma^{SD}_Q \ll \sigma_Q$, where $\sigma^{SD}_Q$ denotes the Slater 
determinant expectation value. If instead, a statistical ensemble of Slater determinants 
is used. Each SD leads to a quantum expectation value $\sigma^{\lambda}_Q $, where $\lambda$ labels the SD under interest.
One can then optimize the statistical ensemble in such a way that $\overline{\sigma^{\lambda}_Q } \simeq \sigma_Q$
where $\overline{~\langle  X \rangle~}$ denotes the statistical average over quantum expectation values. The statistical 
assumption described above provides a practical solution to mimic initial zero point motion in collective space. Then, the 
system evolution is performed by evolving each SD independently from the others, therefore neglecting  
interference between different trajectories. This approach, that can be regarded as the first step towards 
multi-reference TD-EDF, has been shown (i) to 
incorporates the one-body dissipation 
and associated fluctuation mechanism in accordance with the quantal dissipation-fluctuation relation, (ii) to  
give dispersion of one-body observables that is identical to a previous formula
derived from the Balian-V\'en\'eroni variational principle [13,14].   

In practice, initial fluctuations are simulated by an ensemble of
initial single-particle density matrices, each of them associated to a single SD [12]:
\begin{eqnarray}
\label{eq:density}
\rho^\lambda({\mathbf r},{{\mathbf r}}^{\prime},t_0) = \sum\limits_{ij}
\Phi_{i}^\ast({\mathbf r},t_0)\rho_{ij}^\lambda
\Phi_{j}({\mathbf r}^{\prime},t_0),
\end{eqnarray}
where summation $i$ and $j$ is made over a complete set of single-particle states
$\Phi_{i}({\mathbf r},t_0)$, and  $i$ implicitly contains  
spin and isospin quantum numbers. 
Components of density matrices, $\rho_{ij}^\lambda$
are time-independent random Gaussian numbers with mean value
$\overline{\rho_{ij}^\lambda}=\delta_{ij}n_i$
and a variance of the fluctuating part $\delta\rho_{ij}^\lambda$ is 
specified by,
\begin{eqnarray}
\overline{\delta\rho^\lambda_{ij}
\delta\rho_{j'i'}^\lambda }
&=& \frac{1}{2}
\delta_{j{j}'}\delta_{i{i}'}
\left[n_i(1 - n_j) + n_j(1 - n_i)\right].
\label{variance}
\end{eqnarray}
The great advantage of the Stochastic Mean-Field (SMF) theory is that each Slater determinant $\lambda$
evolves independently from each other following the time evolution
of its single-particle wave-functions in its self-consistent mean-field Hamiltonian, denoted by $h(\rho^\lambda)$,
according to
\begin{eqnarray}
\label{eq:spwf}
i\hbar \frac{\partial }{\partial t}\Phi_{i} ({\mathbf r},t;\lambda )
= h(\rho^\lambda )\Phi_{i}({\mathbf r},t;\lambda ).
\end{eqnarray}
with the boundary condition $\Phi_{i} ({\mathbf r},t;\lambda ) = \Phi_{i} ({\mathbf r},t_0)$. Time evolution of mean-values 
and dispersion then express as 
\begin{eqnarray}
Q(t) &=& \sum_{ij} \overline{ \langle \Phi_{i} (t;\lambda) | Q | \Phi_{j} (t;\lambda) \rangle \rho_{ij}^\lambda} \equiv 
\overline{  Q^\lambda (t)}, ~~~~~~
\sigma_Q =  \overline{ ( Q^\lambda (t) - Q(t))^2} ,
\end{eqnarray}
where the average is made over initial conditions.  

\subsection{Application to fusion and transfer reactions}

\begin{figure}[hbtp]
\begin{center}
\includegraphics[width=14cm]{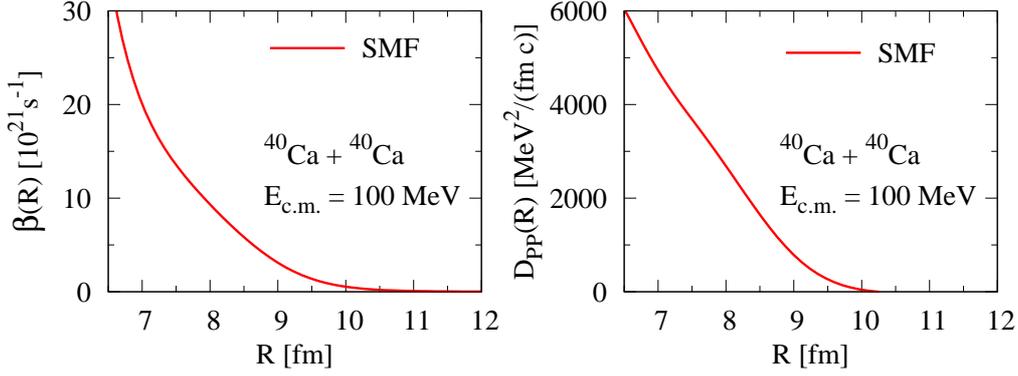}
\caption{(Color online) Evolution of reduced friction (left) and diffusion coefficient (right) as a function of the relative 
distance for the head-on $^{40}$Ca+$^{40}$Ca  collision at center of mass energy $E_{\rm c.m.} = 100$ MeV.}
\label{fig2dubna}
\end{center}
\end{figure}

The powerfulness and applicability of the SMF have been recently 
illustrated in fusion reactions by extending 
the work of ref. [15,16]. Using a macroscopic reduction of the stochastic 
mean-field evolution, central collisions leading to fusion have been mapped to a 
one-dimensional macroscopic Langevin evolution on the relative distance $R$ between the two nuclei
given by [17]:
\begin{eqnarray}
\dot P^{\lambda} = - \partial_R U(R^{\lambda} ) 
- \gamma (R^{\lambda} )\dot {R}^{\lambda} + \xi_P^{\lambda} (t). \label{eq:langevin} 
\end{eqnarray}
$U(R^{\lambda} )$ and $\gamma (R^{\lambda} )$ denote the nuclear+coulomb potential and dissipation 
associated to one-body friction respectively and are already present at the mean-field level [15,16]. $\xi _P^\lambda (t)$ is a 
Gaussian random force acting on the relative
motion reflecting stochasticity in the initial value. This fluctuating part leads to diffusion in collective space 
which can be approximated by $\overline{\xi _P^\lambda (t)\xi _P^\lambda({t}')} \simeq 2\delta(t-{t}')D_{PP}(R)$ where $D_{PP}(R)$ denotes 
the momentum diffusion coefficient. The latter term is nothing but the one that is missing in the original theory and is of primer 
importance to properly describe observables fluctuations. An example of reduced friction $\beta(R)\equiv \gamma(R)/\mu (R)$ and diffusion 
coefficients $D_{PP}(R)$ estimated from the macroscopic reduction of SMF is given in figure \ref{fig2dubna}
for the head-on $^{40}$Ca+$^{40}$Ca  collision.   

\begin{figure}
\label{fig3dubna}
\begin{center}
\includegraphics[width=8.cm, clip]{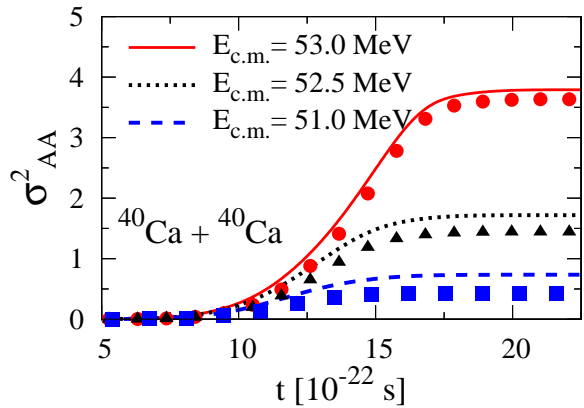}
\caption{(Color online) Evolution of $\sigma_{AA}^2$ calculated in SMF approach for
 $^{40}$Ca${}+^{40}$Ca (top) at different center of mass energies.
Number of exchanged particles is superimposed
by the filled-circles, filled-squares, and filled-triangles
from high to low energies. (Adapted from \cite{WashiPRL})}
\end{center}
\end{figure}

The possibility to estimate transport coefficients associated to fluctuation and dissipation from a fully 
microscopic quantum transport theory is a major breakthrough. However, to be really convincing, one should 
in addition prove that the increase of fluctuations is consistent with experimental observations. To prove that 
SMF can be a predictive framework, we have recently considered transfer reactions [18]. 
Fragment mass distributions deduced from Heavy-Ion reactions have been extensively studied.
It is seen that the dispersion in mass scales approximately with the average number of 
exchanged nucleons. While mean-field properly describes the latter, it miserably fails to account for the 
dispersion. This phenomena is rather well understood in macroscopic models but has not been yet reproduced 
microscopically. To address this issue, we have considered head-on collisions below the Coulomb barrier. 
In that case, nuclei approach, exchange some nucleons and then re-separate. Similarly to the 
relative distance case, a macroscopic reduction onto the projectile (resp. target) mass, denoted by $A^\lambda_P$ (resp. 
$A^\lambda_T$) can be made leading to:
\begin{eqnarray}
\label{eq:langevinaa}
\frac{d}{dt}A_P^\lambda = v(A_P^\lambda, t) + \xi_A^\lambda(t),
\end{eqnarray}
where $v(A_P^\lambda ,t)$ denotes the drift coefficient
for nucleons transfer. Again, the fluctuating term $\xi_A^\lambda (t)$ is linked 
to the diffusion in mass through $\overline{\xi_A^\lambda (t)\xi_A^\lambda ({t}')} = 2\delta(t -{t}')D_{AA}$.
Mass dispersion can then directly be estimated using:
\begin{eqnarray}
\sigma^2_{AA}(t) \simeq 2 \int_0^t D_{AA}(s)ds.
\label{eq:sigma}
\end{eqnarray}
An illustration of $\sigma^2_{AA}(t)$ for  $^{40}$Ca${}+^{40}$Ca  reactions is given in Fig. \ref{fig3dubna}
and compared to the number of exchanged nucleons, denoted by $N_{\rm ex}$. In all cases, both quantities 
are very close from each other and lead to much higher dispersion than the original mean-field.
Indeed, the estimated asymptotic values in the latter case are $0.004$, $0.008$ and $0.008$ from low to high energy
and are much less than the final number of exchanged nucleons that are equal to $0.43$, $1.44$ and $3.63$ respectively.
On opposite, the predicted asymptotic mass dispersions are equal to $0.73$, $1.72$ and $3.79$ and is much closer 
to $N_{ex}$ (see also figure \ref{fig3dubna}).
This numerical test provides a strong support for the validity of the stochastic mean-field approach.

\section{Summary}
Two different approaches have been presented that include correlations beyond mean-field. These approaches 
differ in the strategy and in the types of correlation that are included. The TDDM$^{\rm P}$ formalism is appropriate 
to account for pairing effects in a reasonable numerical time. An illustration on break-up reactions clearly points 
out the importance of such correlations in nuclear reactions. The SMF theory with initial random conditions can be 
seen as the first step towards the inclusion of collective variables zero point motion. Applications to fusion and transfer 
reactions could be of particular interest to treat fluctuations around an average 
mean-field path.      

\bigskip

[1]M. Bender, P.-H. Heenen, and P.-G. Reinhard, Rev. Mod.
Phys. {\bf 75}, 121 (2003).\\
\noindent
[2] C. Simenel, B. Avez, and D. Lacroix, in Lecture notes of the International
Joliot-Curie School,Maubuisson, September 17–22,
2007, arXiv:0806.2714. \\
\noindent
[3] D. Lacroix, T. Duguet, and M. Bender, Phys. Rev. {\bf C79}, 044318 (2009). \\
\noindent
[4]M. Bender, T. Duguet, and D. Lacroix, Phys. Rev. {\bf C79}, 044319 (2009). \\
\noindent
[5]  T. Duguet, M. Bender, K. Bennaceur, D. Lacroix, and
T. Lesinski, Phys. Rev. {\bf C79}, 044320(2009). \\
\noindent
[6] D. Lacroix, S. Ayik and Ph. Chomaz,  Prog. in Part. and Nucl. Phys. {\bf 52}, 497 (2004), 497. \\
\noindent
[7] S. Wang and W. Cassing, Ann. Phys. (N.Y.) {\bf 159}, 328
(1985). \\
\noindent
[8] M. Assi\'e and D. Lacroix, Phys. Rev. Lett. {\bf 102}, 202501 (2009)
; arXiv:0901.0848. \\
\noindent
[9] J. Scarpaci et al., Phys. Lett. {\bf B 428}, 241 (1998). D. Lacroix, J.-A. Scarpaci and Ph. Chomaz, Nucl. Phys. {\bf A658}, 273 (1999).\\
\noindent 
[10] M. Assi\'e et al., Eur. Phys. J. A (2009), {\it in press}. \\
\noindent
[11] Y. Abe, S. Ayik, P.-G. Reinhard and E. Suraud, Phys. Rep. {\bf 275}, 49 (1996). \\
\noindent
[12] S. Ayik, Phys. Lett. {\bf B658}, 174 (2008).\\
\noindent
[13] R. Balian, and M. V\'en\'eroni, Phys. Lett. {\bf B136}, 301 (1984). \\
\noindent
[14] R. Balian, P. Bonche, H. Flocard, and M. V\'en\'eroni,
Nucl. Phys. {\bf A428}, 79c (1984). \\
\noindent
[15] K. Washiyama, D. Lacroix, Phys. Rev. {\bf 78}, 024610 (2008). \\
\noindent
[16] K. Washiyama, D. Lacroix and S. Ayik, Phys. Rev. C {\bf 79}, 024609 (2009). \\
\noindent
[17] S. Ayik, K. Washiyama and D. Lacroix, Phys. Rev. {\bf C79}, 054606, (2009). \\
\noindent
[18] K. Washiyama, S. Ayik and D. Lacroix, arXiv:0609.4498. \\
\end{document}